%% file: main.tex
\documentclass{article}
\usepackage{spconf,amsmath,graphicx}

\usepackage[T1]{fontenc}

\usepackage{amsmath}

\usepackage[utf8]{inputenc} 
\usepackage[T1]{fontenc}    
\usepackage{hyperref}       
\usepackage{url}            
\usepackage{booktabs}       
\usepackage{amsfonts}       
\usepackage{nicefrac}       
\usepackage{microtype}      
\usepackage{soul}
\usepackage[utf8]{inputenc}
\usepackage{graphicx}
\usepackage{amsmath}
\usepackage{booktabs}
\usepackage{outlines}
\usepackage{adjustbox}

\usepackage{subfig}

\usepackage{microtype}
\usepackage{graphicx}
\usepackage{booktabs} 

\input{math_commands.tex}

\usepackage{multirow}
\usepackage{paralist}

\usepackage{booktabs} 
\usepackage{multicol}
\usepackage{diagbox}

\usepackage[ruled,noend,algo2e]{algorithm2e}
\SetCommentSty{mycommfont}

\usepackage{here}

\usepackage{amsmath,amssymb,amsfonts,amsbsy,amsfonts,latexsym}
\usepackage{multirow}

\usepackage{xcolor}
\usepackage{colortbl}

\usepackage{graphicx}

\usepackage[T1]{fontenc}

\usepackage{amsmath}
\usepackage{adjustbox}

\definecolor{colorA}{RGB}{189,201,225}
\definecolor{colorB}{RGB}{103,169,207}
\definecolor{colorC}{RGB}{ 28,144,153}
\definecolor{colorD}{RGB}{  1,108, 89}

\newcolumntype{R}{>{\columncolor{gray!40}}r}
\newcolumntype{L}{>{\columncolor{gray!40}}l}
\newcolumntype{C}{>{\columncolor{gray!40}}c}

\usepackage{tabularx,colortbl,xcolor}
\usepackage{multirow}
\usepackage[normalem]{ulem}
\useunder{\uline}{\ul}{}

\usepackage{enumitem}

\usepackage{xparse}

\DeclareGraphicsExtensions{.pdf,.png}

\DeclareMathOperator*{\argmin}{arg\,min}

\SetKwInput{KwInput}{Input}
\SetKwInput{KwRequire}{Require}

\usepackage{longtable}
\usepackage{pgfplots}

\usepackage{lipsum}

\usepackage{amssymb}
\usepackage{pifont}

\usepackage{adjustbox}

\input macros.tex

\title{Integer-only Zero-shot Quantization for Efficient Speech Recognition}
\usepackage{spconf}
\makeatletter
\def\@name{
\emph{
Sehoon Kim, 
Amir Gholami,
Zhewei Yao,
Nicholas Lee,
Patrick Wang,
}
\\ \emph{
Aniruddha Nrusimha, 
Bohan Zhai, 
Tianren Gao, 
Michael W. Mahoney,
Kurt Keutzer
}
\\
}
\makeatother
\address{University of California, Berkeley}
\begin{document}
%
\maketitle
\begin{abstract}
End-to-end neural network models achieve improved performance on various automatic speech recognition (ASR) tasks.
However, these models perform poorly on edge hardware due to large memory and computation requirements.
While quantizing model weights and/or activations to low-precision can be a promising solution, previous research on quantizing ASR models is limited. 
In particular, the previous approaches use floating-point arithmetic during inference and thus they cannot fully exploit efficient integer processing units.
Moreover, they require training and/or validation data during quantization,
which may not be available due to security or privacy concerns. 
To address these limitations, we propose an integer-only, zero-shot quantization scheme for ASR models.
In particular, we generate synthetic data whose runtime statistics resemble the real data, and we use it to calibrate models during quantization. 
We apply our method to quantize QuartzNet, Jasper, and Conformer and show negligible WER degradation as compared to the full-precision baseline models, 
even without using any data.
Moreover, we achieve up to 2.35$\times$ speedup on a T4 GPU and $4\times$ compression rate,
with a modest WER degradation of <1\%  with INT8 quantization.
\end{abstract}

\vspace{-2mm}

\section{Introduction}
\label{sec:intro}
\vspace{-1mm}

End-to-end neural network models have achieved  state-of-the-art results in various automatic speech recognition (ASR) tasks~\cite{li2019jasper, gulati2020conformer, han2020contextnet, zhang2020transformer, kriman2020quartznet}.
However, these accuracy improvements come with increasingly large model sizes.
For instance, the highest performing versions of Jasper~\cite{li2019jasper}, Conformer~\cite{gulati2020conformer}, ContextNet~\cite{han2020contextnet}, and Transformer-Transducer~\cite{zhang2020transformer} contain 333M, 118.8M, 112.7M, and 139M parameters. 
This has made real-time deployment of these models challenging at the edge.

One promising solution to tackle this challenge is to quantize model weights and/or activations to low-precision.
This provides multiple benefits.
First, quantization reduces the memory footprint by storing weights and/or activations in low-precision.
For instance, INT8 uses 4$\times$ less memory relative to FP32.
Second, quantization accelerates execution and decreases power consumption by using specialized hardware for low-precision arithmetic~\cite{jacob2018quantization, yao2020hawqv3, kim2021bert}.
Following its success on many computer vision~\cite{jacob2018quantization, yao2020hawqv3} and natural language processing tasks~\cite{kim2021bert, zafrir2019q8bert, shen2020q}, there have been attempts to apply quantization to ASR models~\cite{yao2020int8, nguyen2020quantization, prasad2020quantization, bie2019simplified}.
However, prior quantization work lacks two important features:
integer-only quantization and zero-shot quantization.

\textit{Integer-only quantization}~\cite{jacob2018quantization, yao2020hawqv3, kim2021bert} is a quantization scheme where \emph{all} operations (e.g., convolution and matrix multiplication) are performed using low-precision integer arithmetic.
To the best of our knowledge, all the prior quantization methods for ASR models use \textit{simulated quantization}, where all or part of operations are performed with floating point arithmetic.
For instance, \cite{prasad2020quantization} only performs convolution with integer arithmetic and leaves ReLU and Batch Normalization (BatchNorm) as floating point operations.
Compared to simulated quantization, integer-only quantization can decrease latency and power consumption by fully utilizing efficient integer processing units. 
More importantly, it also allows deployment on popular and highly optimized edge processors designed for embedded, mobile, or IoT devices that often do not support floating point arithmetic.
ARM Cortex-M~\cite{armcortexm}, GreenWaves GAP-9~\cite{flamand2018gap}, and Google's Edge TPU~\cite{edgetpu} are some examples of edge processors without dedicated floating point units.

\textit{Zero-shot quantization}~\cite{chen2019data, haroush2020knowledge, cai2020zeroq, choi2020data, xu2020generative,he2021generative} is a quantization scheme that does not require any training or validation data.
This is important because datasets are not always available, especially for ASR use cases such as smart speakers or healthcare, where privacy and security are concerns.
However, prior work is based on either quantization-aware training~\cite{yao2020int8, nguyen2020quantization, prasad2020quantization, bie2019simplified} or post-training quantization~\cite{prasad2020quantization, bie2019simplified},
both of which require access to training and/or validation data to finetune or calibrate the quantized models.
As such, these methods cannot be applied if no data is~available.

We propose  to address the aforementioned limitations of the prior quantization work for ASR models. In particular:
\begin{itemize}[leftmargin=*]
    \vspace{-2mm}
    \item We develop an integer-only, zero-shot quantization scheme for ASR models, which allows efficient deployment on integer  processing  units and achieves negligible WER degradation even without any access to  training and/or validation data (\sref{subsec:integer_only} and \sref{subsec:zsq}).
    \vspace{-2.5mm}
    \item We apply our method to QuartzNet-15x5~\cite{kriman2020quartznet}, JasperDR-10x5~\cite{li2019jasper}, and Conformer-Large~\cite{gulati2020conformer} and evaluate their word-to-error-rate (WER) on the Librispeech benchmark~\cite{panayotov2015librispeech}.
    With 8-bit quantization for weights and activations, we achieve negligible WER degradation of up to 0.29\%, 0.08\%, and 0.87\% for each model, respectively (\sref{sec:accuracy}).
    
    \vspace{-2mm}
    \item 
     We deploy INT8-only QuartzNet on a T4 GPU, and show up to $2.35\times$ speedup compared to its FP32 counterpart (\sref{sec:latency}).
\end{itemize}

\vspace{-3mm}
\section{Methodology}
\label{sec:methods}
\vspace{-1mm}
\subsection{Basic Quantization Method}
\label{sec:basic}
In this work, we use \textit{uniform symmetric quantization}.  This method uniformly maps a real number $x$ to an integer value $q$:
\begin{equation}
\label{eq:uniform_quantization}
q = \mathrm{Q}(x, b, S) = \mathrm{Int}\bigg(\frac{\mathrm{clip}(x, -\alpha, \alpha)}{S}\bigg),
\end{equation}  
where $q \in [-2^{b-1}, 2^{b-1} - 1]$, $b$ is the quantization bit-width,
$\mathrm{Q}$ is the quantization operator, $\mathrm{Int}$ is the round operation, $\mathrm{clip}$ is the truncation function with the clipping parameter $\alpha$, and $S$ is the scaling factor defined as $\alpha / (2^{b-1} - 1)$.
The reverse mapping from the quantized values $q$ to the real values (aka dequantization) is
$\tilde{x} = \mathrm{DQ}(q, S) = Sq \approx x$, where $\mathrm{DQ}$ denotes the dequantization operator. 

Determining the clipping range $[-\alpha, \alpha]$ that best represents the range of the input $x$ is a primary challenge of quantization. 
One popular choice in practice is to use the minimum and maximum values of the input $x$, i.e., $\alpha ={\max(|x_{max}|, |x_{min}|)}$.
However, this approach is susceptible to outliers. 
An unnecessarily large clipping range due to a few outliers increases the rounding error of quantized values within the range.
One way to alleviate this is to use percentile values, e.g., 99$\%$ smallest/largest, instead of the min/max values.

In \textit{dynamic quantization}, clipping ranges are computed during inference. 
However, calculating input statistics (e.g., min, max, and percentile) can be  costly in real-time, and requires floating point operations that would prevent us from doing integer-only quantization.
Therefore, in this work, we only consider \textit{static quantization} where we pre-compute the clipping ranges and fix them during inference as in~\cite{jacob2018quantization, yao2020hawqv3, kim2021bert}.
It is straightforward to pre-compute the ranges for weights as they have fixed values during inference.
However, activations vary across different inputs, and therefore their ranges also vary.
One popular method to address this is \textit{calibration},
which runs a series of training data to compute the typical range of activations.
Later in~\sref{subsec:zsq}, we extend this idea and show how to calibrate \textit{without} training data.
We refer the interested readers to~\cite{gholami2021a} for more details in quantization methods.

\vspace{-2mm}
\subsection{Integer-only Quantization}
\label{subsec:integer_only}
Integer-only quantization~\cite{jacob2018quantization, yao2020hawqv3, kim2021bert} not only represents the model weights and activations with low-precision integer values, but it also carries out the entire inference with integer arithmetic.
Broadly speaking, the core of integer-only quantization is the linear property of the operations.
For instance, let $q_W$ and $q_x$ denote the quantized values for weight $\tilde{W}$ and activation $\tilde{x}$, respectively.
That is, $\tilde{W} = S_W q_W$ and $\tilde{x} = S_x q_x$ where $S_W$ and $S_x$ are the scaling factors for $\tilde{W}$ and $\tilde{x}$ (\sref{sec:basic}).
Then, $\mathrm{Conv}(\tilde{W}, \tilde{x}) = \mathrm{Conv}(S_W q_W, S_x q_x)$ 
is equivalent to $S_W S_x \mathrm{Conv}(q_W, q_x)$ due to the linear property.
Therefore, we can apply \textit{integer} convolution directly to the quantized values $q_W$ and $q_x$ without having to dequantize them into $\tilde{W}$ and $\tilde{x}$ and apply floating point convolution~\cite{jacob2018quantization,gholami2021a}.
This property holds true for convolution, matrix multiplication, and ReLU activation, which are the basic building blocks for neural architectures.
In addition, BatchNorms are folded into the preceding convolution layers~\cite{jacob2018quantization}, and non-linear activations (e.g., Sigmoid, Softmax and Swish in Conformer) are approximated with 2nd-order polynomials following the procedure in~\cite{kim2021bert} to avoid their floating point execution.

\vspace{-2mm}
\subsection{Zero-shot Quantization}
\label{subsec:zsq}
Quantization generally requires training data (1) to pre-compute the clipping ranges for calibration (\eref{eq:uniform_quantization}) and/or (2) to finetune the quantized model 
to recover accuracy degradation from quantization.
However, in many cases, pre-trained models are distributed without the data they were trained on due to proprietary, privacy, and security concerns. 
As such, multiple zero-shot quantization schemes have been proposed, mostly in the field of computer vision, to allow quantization in such cases.
A widely adopted method is to generate synthetic image data that can be used in place of the real training data for calibration and/or finetuning~ \cite{chen2019data, haroush2020knowledge, cai2020zeroq, choi2020data, xu2020generative,he2021generative}.

Similar to~\cite{haroush2020knowledge, choi2020data}, we generate synthetic data with a similar data distribution to the training data.
In particular, we generate the data to match the runtime statistics (i.e., hidden activation distributions) of the training data using the running means and variances stored in BatchNorm layers.
Note that the running means and variances capture the runtime statistics of the training data.
Let $x$ denote the synthetic data, 
and $\mu_i$ and $\sigma_i$ the mean and variance of the activation from the $i$-th BatchNorm with $x$ as an input.
Additionally, let $\hat{\mu}_i$ and $\hat{\sigma}_i$ denote the running mean and variance of the BatchNorm.
Then, we aim to minimize the loss defined by the Kullback-Leibler (KL) divergence between $N(\hat\mu_i, \hat\sigma_i)$ and $N(\mu_i, \sigma_i)$, assuming that the hidden activations follow the Gaussian distribution:
\begin{equation}
\small
\begin{split}
\label{eq:kl}
    x^* &= \argmin_{x} \sum_{i} KL(N(\hat\mu_i, \hat\sigma_i) || N(\mu_i, \sigma_i)) \\
    &= \argmin_{x} \sum_{i} \log{\frac{\sigma_i}{\hat\sigma_i}}
       - \frac{1}{2}\left(1 - \frac{{\hat\sigma_i}^2 + (\hat\mu_i - \mu_i)^2}{\sigma_i^2}\right) .
\end{split} 
\end{equation}  

\begin{figure}[]
\centering{
\centerline{
  \includegraphics[width=0.45\textwidth]{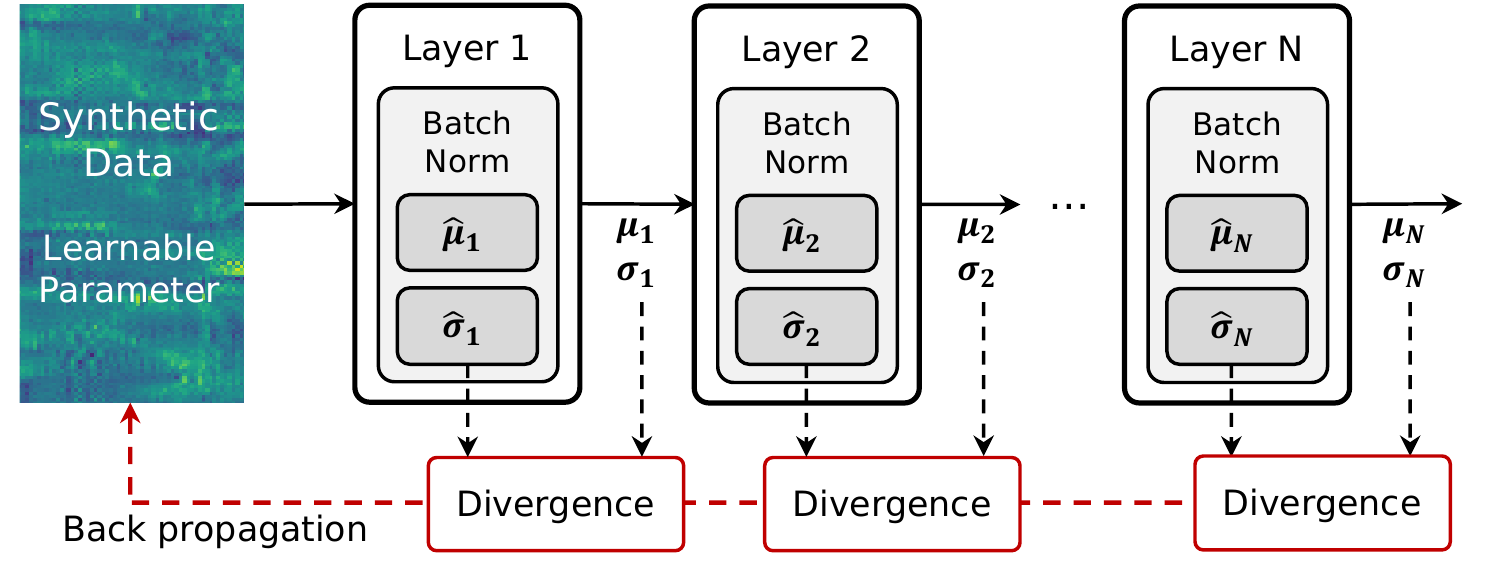}
  }
  \vspace{-2mm}
  \caption{
  End-to-end process of synthetic data (i.e., mel spectrogram) generation for zero-shot quantization. 
  Input data is trained with standard backpropagation in a direction that matches the hidden activation statistics of the real data.
  }
  \label{fig:zsq}
  }
  \vspace{-5.5mm}
\end{figure}

In computer vision tasks, input $x$ is often regarded as a learnable parameter and is trained with standard backpropagation to reduce the defined loss~\cite{haroush2020knowledge, cai2020zeroq}.
However, it is not straightforward to apply this directly to speech data, as we are dealing with time varying audio signals rather than static images.
As such, we instead propose to generate synthetic mel spectograms.
Mel spectrogram is a widely used preprocessed representation of audio signals~\cite{li2019jasper, gulati2020conformer, han2020contextnet, kriman2020quartznet, park2019specaugment}.
Also, its data representation is more suitable for synthetic data generation as it can be regarded as a 1-dimensional image.
Specifically, our zero-shot quantization scheme consists of two steps.
\begin{itemize}[leftmargin=*]
    \vspace{-1mm}
    \item \textbf{Synthetic data generation:} 
    A batch of 1-dimensional arrays is randomly initialized and set as a learnable parameter.
    It is then fed into the full-precision (i.e., non-quantized) model to compute the loss in \eref{eq:kl}.
    This minimizes the KL divergence between two distributions, one from the real training data and the other from the synthetic data, for each BatchNorm activation.
    Using backpropagation, the input is iteratively trained using gradient descent to minimize the loss.
    \fref{fig:zsq} illustrates the end-to-end process.
    
    \vspace{-2mm}
    \item \textbf{Calibration:} 
    The synthetic data is fed into the target model in place of the real mel spectrograms to determine the clipping ranges for all activations.
    Then the model is quantized according to~\eref{eq:uniform_quantization}.
\end{itemize}
\vspace{-1mm}
The benefit of this approach is that we can use this synthetic data to perform quantization, and avoid the
need to get direct access to the training data which may
not be available.

\vspace{-3mm}
\section{Results}
\label{sec:results}
\vspace{-2mm}

In this section, we first demonstrate in~\sref{sec:accuracy} that the quantized models can match the accuracy of full-precision baselines.
We then evaluate the latency speedup of the quantized models on real hardware in~\sref{sec:latency}.
As baseline models, we select QuartzNet15x5~\cite{kriman2020quartznet} and JasperDR-10x5~\cite{li2019jasper} as CNN-based architectures, and Conformer-Large~\cite{gulati2020conformer} as a more complex architecture with multi-head attentions and non-linear activations such as Sigmoid, Softmax, and Swish.
Such non-linear activations can be computed with integer arithmetic by approximating them with 2nd-order polynomials as described in~\cite{kim2021bert}.
We note that our quantization method is not a model-specific solution, and can be applied to many other ASR models.


\vspace{-2mm}
\subsection{Accuracy Evaluation}
\label{sec:accuracy}


\begin{table}[t]
\caption{ 
WER of the quantized QuartzNet15x5, JasperDR-10x5, and Conformer-Large with different bit-width settings, evaluated on LibriSpeech.
W and A are the bit-width for weights and activations.
Note that 32/32 indicates full-precision baseline.
We also include the model size and BOPs (for 10-second input with sampling rate 16K) of each model.
}
\vspace{-1.5mm}
\label{tab:accuracy}
\centering
\centerline{
\footnotesize{
    \setlength{\tabcolsep}{4.5pt}{
      \begin{tabular}[t]{c|c|cccc|cc}
        \toprule
        \ha  \multirow{2}{*}{Model}  & \multirow{2}{*}{W/A} & \multicolumn{2}{c}{dev}  & \multicolumn{2}{c|}{test}  & \multicolumn{1}{c}{Size} & \multicolumn{1}{c}{ BOPs} \\ 
        \  &    &  clean & other & clean & other & (MB)  & (T) \\

        \midrule
        \ha  QuartzNet &  32/32  &  3.80 & 10.05 & 3.82 & 10.08  & 73.81 & 9.48 \\
        \hbb    & 8/8 &   3.92& 10.28	& 4.04 & 10.37 & 18.45 & 0.61 \\
        \hb     &  6/8  &   4.16 & 10.83 & 4.32 & 11.06  & 13.84 & 0.46\\
        \midrule
        \ha  Jasper &  32/32  & 3.47 & 10.40 & 3.68 & 10.49  & 1208 & 171 \\
        \hbb       & 8/8 & 3.47 & 10.49 & 3.68 & 10.57  & 302.0 & 10.7\\
        \hb     &  6/8  & 3.52 & 10.59 & 3.73 & 10.68 & 226.5 & 8.00 \\
        \midrule
        \ha  Conformer &  32/32  & 2.47 & 6.04 & 2.78 & 6.19 & 494.8 & 44.0 \\
        \hbb       & 8/8          & 2.75 & 6.95 & 3.06 & 7.06 & 123.7 & 2.75 \\
        \hb     &  6/8  &         3.63 & 8.21 & 4.03 & 8.48 & 92.78 & 2.06 \\
        \bottomrule
        
        \end{tabular} 
        }
        }
}

\vspace{-4mm}
\end{table}


For accuracy evaluation, we use the NeMo~\cite{nemo} implementation of the pre-trained QuartzNet-15x5, JasperDR-10x5, and Conformer-Large for the full-precision baseline models. 
The synthetic data is initialized with the uniform distribution from $[-0.3, 0.3]$, 
and is trained using the Adam optimizer~\cite{kingma2014adam} with batch size 8, $\beta_1=0.9$, $\beta_2=0.999$, and learning rate of $\{$0.03, 0.04, 0.05, 0.06$\}$ for $\{$200, 250$\}$ iterations.
We generate 20 such batches and use them for calibration.
Figure~\ref{fig:examples} contains some examples of the generated synthetic data.
For calibration, we use the min/max values for clipping.
We quantize the baseline models with W8A8 (i.e., 8-bit weights and activations) and W6A8,
and do not finetune the models after quantization.
All the reported numbers are averaged over 4 different runs (i.e., synthetic data generation and calibration).

Both the baseline and quantized models are evaluated on the widely used LibriSpeech~\cite{panayotov2015librispeech} benchmark.
Table~\ref{tab:accuracy} compares the WER of the baseline and quantized models with different bit-width settings on dev-clean/other and test-clean/other datasets, where hyperparameters are tuned on the dev sets.
We also include the model size and BOPs (bit-operations) of each model for comparison.
BOP is the total number of bit operations and is a hardware-agnostic proxy to model complexity~\cite{van2020bayesian}.
For QuartzNet, the W8A8 setting results in a modest WER degradation of 0.23\% and 0.29\% on the test-clean/other datasets with 4$\times$ reduction of the model size 
as compared to the full-precision baseline.
The W6A8 setting further reduces the model size to $\sim$5$\times$ of the baseline only within 0.50\% and 0.97\% WER degradation on the test sets.
On Jasper, the quantized model exhibits negligible WER increase of up to 0.08$\%$ and 0.19$\%$ on the test sets with the W8A8 and W8A6 settings, respectively.
Despite the complexity of the Conformer architecture, INT8-only version only exhibits 0.87\% WER degradation.
Note that all the reported values are \textit{without} any finetuning after quantization.

\vspace{-2.5mm}
\subsection{Latency Evaluation}
\label{sec:latency}
We evaluate the latency speedup of quantized models by direct deployment of the quantized QuartzNet on a Tesla T4 GPU with the Turing Tensor Cores that support accelerated INT8 execution. 
We use NVIDIA's TensorRT library~\cite{tensorrt} for model deployment.
All the experiments were conducted on Google Cloud Platform virtual machine with a single Tesla T4 GPU, CUDA 11.1, cuDNN 8.0, and TensorRT 7.2.1.
Although we select T4 GPU as our target device due to its extensive software support~\cite{tensorrt}, 
we should highlight that our approach is not specific to GPUs, and one could 
also deploy the quantized models to other processors.
For evaluation, we use QuartzNet-5x5, 10x5, and 15x5, which are the small, medium, and large variants of the QuartzNet family~\cite{kriman2020quartznet}.
We measure the inference latency using a 10-second audio input with a sampling rate 16K, and test the W8A8 (i.e., INT8-only) setting for the quantized model.

\begin{figure}[]
\centering{
\centerline{
  \includegraphics[width=0.45\textwidth]{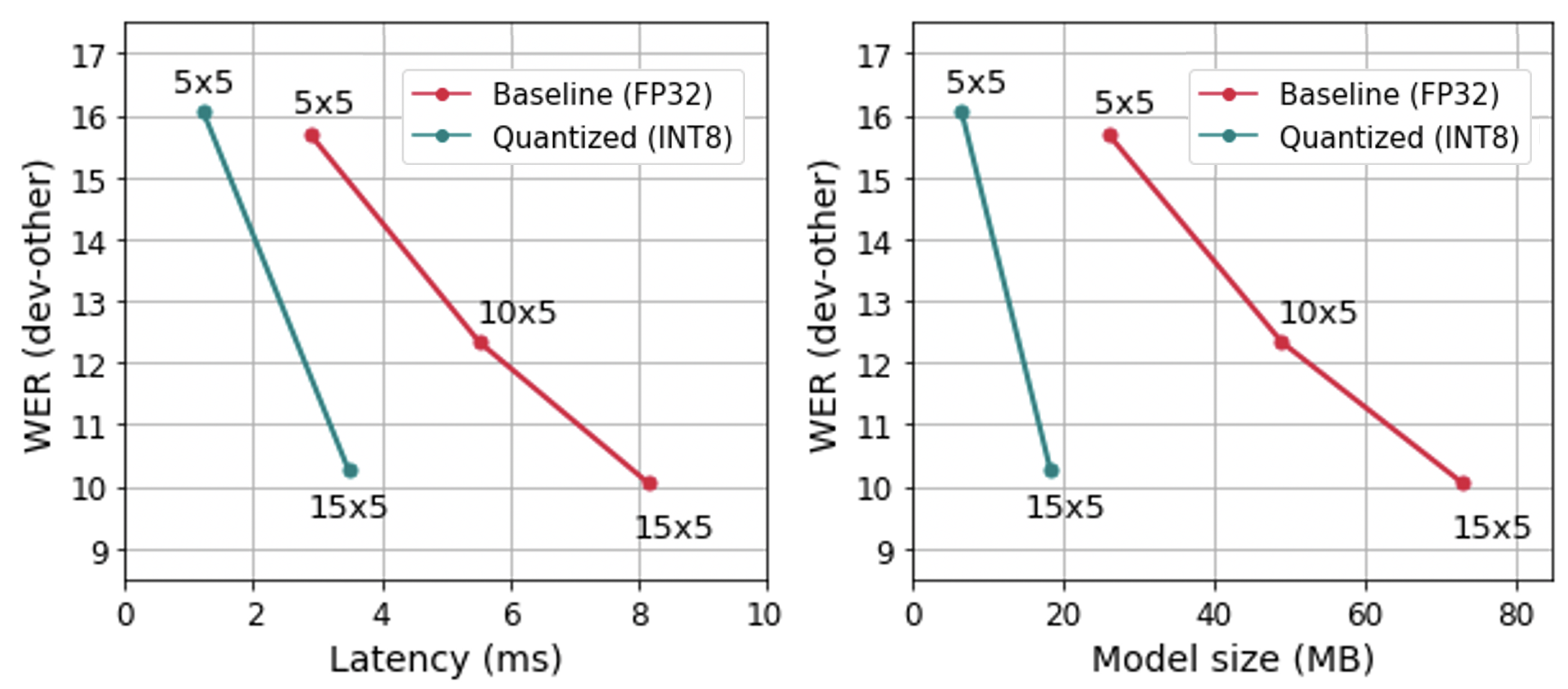}
  }
  \vspace*{-2mm}
  \caption{
    Latency (Left) and model size (Right) of the quantized and baseline  QuartzNets with different sizes.
    Latency is measured with 10-second audio input with sampling rate 16K.
  }
  \label{fig:latency}
  }
  \vspace{-5mm}
\end{figure}

Figure~\ref{fig:latency} compares the latency and model size of the baseline and quantized QuartzNet models with different size configurations.
As shown in the plot, INT8-only QuartzNet15x5 achieves 2.35$\times$ speedup and 4$\times$ compression rate as compared to the FP32 baseline.
Also, the quantized QuartzNet15x5 matches the WER of the largest model, while only needing the computation and memory requirements of the smallest model.
This is an important observation for practitioners who wants to design an efficient model, as the alternative approach of using a full precision but shallower model can lead to a  sub-optimal solution compared to quantization.

\vspace{-2mm}
\section{Discussion}
\vspace{-1mm}
\subsection{Ablation studies}

Here, we show the effectiveness of the zero-shot quantization scheme.
To do so, we quantize QuartzNet15x5 using a set of random calibration data from the uniform distribution in $[-3, 3]$.
We observe significant WER degradation when we calibrate the model with random data instead of the synthetic data of our zero-shot quantization scheme.
With random data, the test-other WER degrades by 2.47\% and 4.02\% for W8A8 and W6A8, respectively, which are noticeably higher
than the corresponding values with the synthetic data as shown in~\tref{tab:ablation}.
The results confirm the need to generate synthetic data that well represents the actual data used in training.

\begin{table}[!t]
\caption{  
WER of the quantized QuartzNet15x5 calibrated with random data and the synthetic data.
W and A are the bit-width for weights and activations, respectively.
Calibration with the synthetic data consistently outperforms calibration with random data for all bit-width settings.
}
\vspace{-2mm}
\label{tab:ablation}

    \centering

\footnotesize{
    \setlength{\tabcolsep}{4.5pt}{
      \begin{tabular}[t]{c|c|cccc|cc}
        \toprule
        \ha  \multirow{2}{*}{Method}  & \multirow{2}{*}{W/A} & \multicolumn{2}{c}{dev}  & \multicolumn{2}{c|}{test}  & \multicolumn{1}{c}{Size} & \multicolumn{1}{c}{ BOPs} \\ 
        \  &    &  clean & other & clean & other & (MB)  & (T) \\
        \midrule
        \ha  Baseline &  32/32  &  3.80 & 10.05 & 3.82 & 10.08  & 73.81 & 9.48 \\
        \midrule
        \ha  Random    & 8/8 &   5.39 & 12.20 & 5.56 & 12.55 & 18.45 & 0.61 \\
        \hb  Synthetic    & 8/8 &   3.92& 10.28	& 4.04 & 10.37 & 18.45 & 0.61 \\
        \midrule
        \ha    Random &  6/8  &   6.26 & 13.57 & 6.56 & 14.10   & 13.84 & 0.46\\
        \hb   Synthetic  &  6/8  &   4.16 & 10.83 & 4.32 & 11.06  & 13.84 & 0.46\\
        \bottomrule
        \end{tabular} 
        }
        }
\end{table}


    \vspace{-2mm}
\subsection{Analysis of Synthetic Data Generation}

\begin{figure}[]
\centering{
\centerline{
  \includegraphics[width=0.42\textwidth]{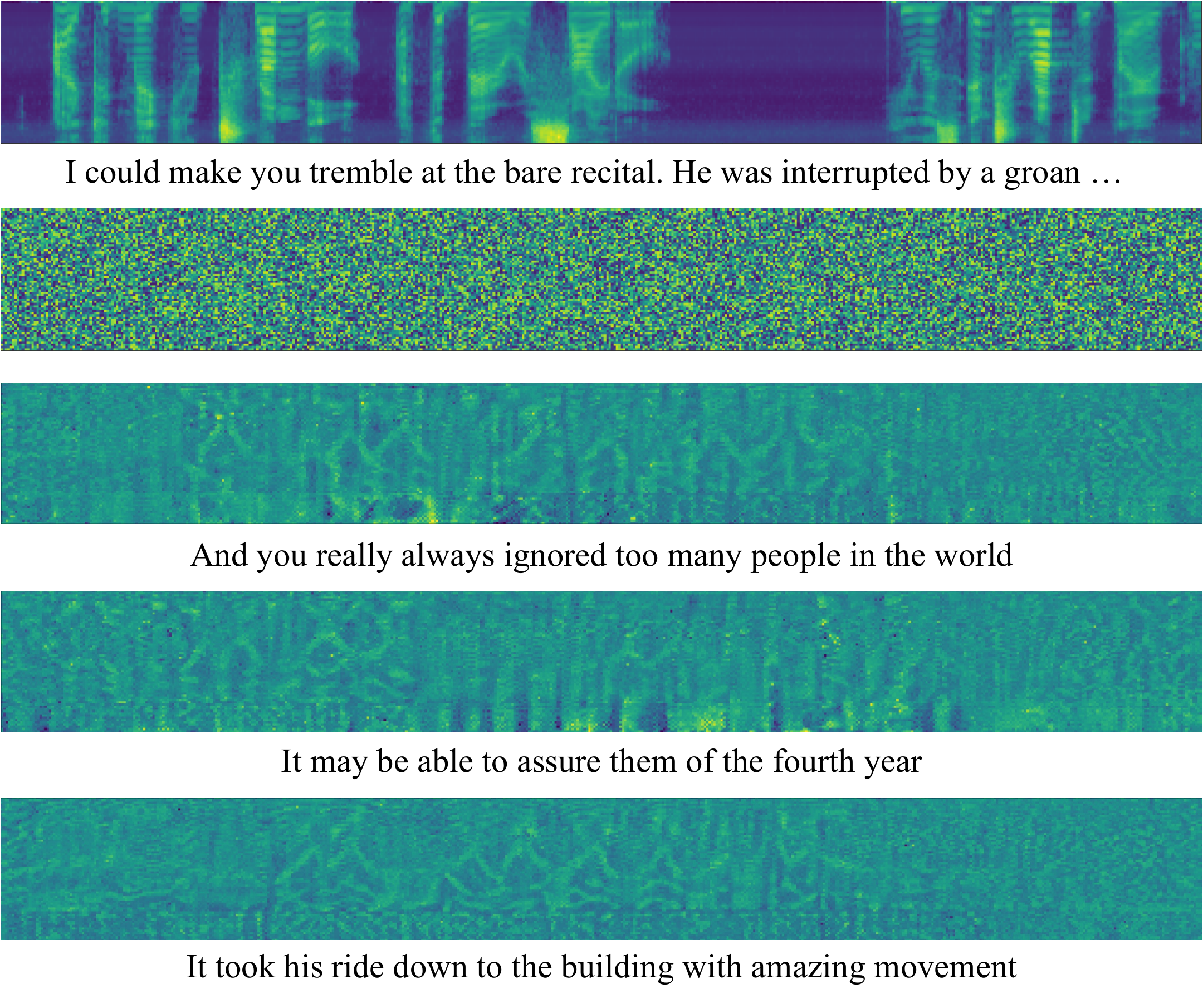}
  }
  \vspace*{-2mm}
  \caption{
    Comparison of the real (1st), random (2nd), and synthetic (3rd-5th) mel spectrograms.
    Below each synthetic mel spectrogram is the text that it decodes to.
  }
  \label{fig:examples}
  }
  \vspace{-6mm}
\end{figure}

In this section, we briefly analyze the synthetic data,
which we start with a randomly initialized array (second row of~\fref{fig:examples}) and solve~\eref{eq:kl} to generate synthetic mel spectrograms.
We show two such examples in the last three rows of~\fref{fig:examples}.
First, note that the synthetic data can capture the local patterns observed in real mel spectrograms.
Second, we observe that if the synthetic data is fed into QuartzNet followed by a 3-gram language model, it can be decoded into meaningful words and sentences, instead of a sequence of random characters.
Examples of the decoded sentences are captioned below the corresponding synthetic data in~\fref{fig:examples}.
This is a further evidence that the synthetic data is close to the actual training data, and it explains why it enables good quantization results.

\vspace{-2.5mm}
\section{Conclusion}
\label{sec:conclusion}
\vspace{-2mm}

In this work, we propose a zero-shot quantization scheme for ASR models that only uses integer-only computation.
Our proposed method entirely removes floating point operations from inference,
thereby allowing efficient model deployment on fast and power-efficient integer processing units.
In addition, it does not require any training and/or validation data, as it calibrates models with synthetic data that resembles the real data.
Our quantization scheme exhibits a large compression rate of $4\times$ with small WER degradation of 
0.29\%, 0.08\%, and 0.87\% on QuartzNet, Jasper, and Conformer for INT8-only quantization.
Furthermore, we achieve 2.35$\times$ speedup by directly deploying INT8-only QuartzNet on a T4 GPU.

\vfill\pagebreak

\small{
\bibliographystyle{IEEEbib}
\bibliography{strings,refs}
}
\end{document}

%% file: math_commands.tex
\usepackage{amsmath,amsfonts,bm}

\usepackage{xcolor}
\usepackage{colortbl}

\def\eqref#1{equation~\ref{#1}}

\DeclareMathAlphabet{\mathsfit}{\encodingdefault}{\sfdefault}{m}{sl}
\SetMathAlphabet{\mathsfit}{bold}{\encodingdefault}{\sfdefault}{bx}{n}



%% file: macros.tex
\usepackage{siunitx}

\newcommand{\commentout}[1]{}

\newcommand\eref{Equation~\ref}
\newcommand\fref{Figure~\ref}
\newcommand\tref{Table~\ref}
\newcommand\sref{\S~\ref}

\newcommand\ha{ \rowcolor{orange!0}}
\newcommand\hb{ \rowcolor{black!10}}
\newcommand\hbb{ \rowcolor{black!5}}

\usepackage{amsthm}

\usepackage{amsthm}

\newtheorem{mytheorem}{Theorem}
\newtheorem{assumption}[mytheorem]{Assumption}
\newtheorem{lemma}[mytheorem]{Lemma}
\newtheorem{remk}[mytheorem]{Remark}
